\documentclass[10pt,superscriptaddress,twocolumn,amsmath,amssymb,aps,prl,showpacs]{revtex4}
\usepackage{mathrsfs}
\usepackage{graphicx}
\usepackage{dcolumn}
\usepackage{bm}
\usepackage{amssymb}
\usepackage{amsmath}
\usepackage{paralist}
\usepackage{pdfpages}
\usepackage{amsmath,amssymb,mathrsfs}

\def\be{\begin{equation}}
\def\ee{\end{equation}}
\def\bea{\begin{eqnarray}}
\def\eea{\end{eqnarray}}

\begin{document}

\title{ FFLO correlation  and  free  fluids  in the one-dimensional  attractive Hubbard model
}

\author{Song Cheng}
\affiliation{State Key Laboratory of Magnetic Resonance and Atomic and Molecular Physics,
Wuhan Institute of Physics and Mathematics, Chinese Academy of Sciences, Wuhan 430071, China}
\affiliation{University of Chinese Academy of Sciences, Beijing 100049, China.}
\affiliation{Department of Theoretical Physics, Research School of Physics and Engineering,
Australian National University, Canberra ACT 0200, Australia}

\author{Yi-Cong Yu}
\affiliation{State Key Laboratory of Magnetic Resonance and Atomic and Molecular Physics,
Wuhan Institute of Physics and Mathematics, Chinese Academy of Sciences, Wuhan 430071, China}
\affiliation{University of Chinese Academy of Sciences, Beijing 100049, China.}

\author{M. T. Batchelor}
\affiliation{Centre for Modern Physics, Chongqing University, Chongqing 400044, China}
\affiliation{Department of Theoretical Physics,
Research School of Physics and Engineering,
Australian National University, Canberra ACT 0200, Australia}
\affiliation{Mathematical Sciences Institute, Australian
National University, Canberra ACT 0200, Australia}

\author{Xi-Wen Guan}
\email[]{xiwen.guan@anu.edu.au}
\affiliation{State Key Laboratory of Magnetic Resonance and Atomic and Molecular Physics,
Wuhan Institute of Physics and Mathematics, Chinese Academy of Sciences, Wuhan 430071, China}
\affiliation{Department of Theoretical Physics,
Research School of Physics and Engineering,
Australian National University, Canberra ACT 0200, Australia}

\affiliation{Center for Cold Atom Physics, Chinese Academy of Sciences, Wuhan 430071, China}

\date{\today}

\pacs{71.10.Fd, 75.40.Cx,02.30.Ik}

\begin{abstract}
 In this Rapid Communication  we show  that low energy macroscopic properties of the one-dimensional (1D)
 attractive Hubbard model exhibit two fluids of bound pairs and of unpaired fermions.
 Using the thermodynamic  Bethe ansatz equations  of the model, we  first  determine the low temperature phase diagram  and analytically calculate
 the Fulde-Ferrell-Larkin-Ovchinnikov (FFLO) pairing correlation function for  the partially-polarized phase.
We  then  show  that for such a FFLO-like  state  in the low density regime the  effective chemical potentials of bound pairs and unpaired fermions
behave like  two free fluids.
 Consequently, the susceptibility, compressibility and specific heat  obey  simple  additivity rules, indicating
 the `free' particle  nature of interacting fermions on a 1D lattice.
 In contrast to the  continuum Fermi gases,   the correlation critical exponents and  thermodynamics of the
 attractive Hubbard  model  essentially depend on two lattice  interacting parameters.
Finally, we  study scaling functions, the Wilson ratio  and susceptibility  which  provide universal  macroscopic
properties and dimensionless constants   of interacting fermions at low energy.

\end{abstract}

\maketitle

\noindent
The notion of Landau  quasiparticles  gives rise to the  Fermi liquid  theory  successfully used for describing
properties of a large variety of systems, such as Fermi liquid ${}^3\textrm{He}$ and electrons in metals \cite{Hew97}.
In contrast, it is generally accepted that Fermi liquid theory is not applicable in 1D,
where  the description of the low-energy physics of strongly correlated electrons, spins, bosonic and fermionic atoms relies
on the  Tomonaga-Luttinger liquid (TLL) theory \cite{Gia04}.
Such an understanding of the TLL  in 1D is based on collective
excitations which are significantly different from  Landau  quasiparticles in higher dimensions.
However, concerning macroscopic properties, there are many universal properties/quantities
which are common for both 2D/3D and 1D systems \cite{Wan98,Carmelo:1992,Yu:2016,Shaginyan:2016}.

The  1D repulsive Fermi-Hubbard model describing interacting fermions on a lattice provides a paradigm for understanding many-body physics,
including  spin-charge separation, fractional excitations, quantum dynamics of spinons,  a Mott insulating  phase and magnetism \cite{Ess05}.
Very recently, ultracold atoms trapped in   optical lattices \cite{Hart:2015,Lawrence:2016,Singha:2016,Parsons:2016,Zhang:2015}
offer promising opportunities  to test such fundamental concepts \cite{Boll:2016}.
In contrast, the 1D attractive Fermi-Hubbard model \cite{Krivnov:1975,Bogoliubov:1988,Lee:1988,Penc,Woynarovich:1983,Sacramento:1994,Essler:1994}
is a notoriously  difficult problem due to the complicated  bound states of multi-particles and multi-spins  on lattices.
Despite there being a mapping by Shiba transformations between the repulsive and attractive regions of the Hubbard model \cite{Ess05},
 such a mapping cannot be used for a study of  the low energy themodynamics  of the attractive Hubbard model
 due to the different  cut-off  processes in terms of  such multi-spin and multi-charge bound states.
Of central importance to  this attractive Hubbard model  is the understanding  of quantum  correlations of charge  bound states,  for example,
the  Fulde-Ferrell-Larkin-Ovchinnikov (FFLO) like  pairing \cite{Fulde:1964,Larkin:1965}  on a 1D lattice \cite{Yang:2001,Tezuka:2008,Feiguin:2007}.
In the expansion dynamics of the FFLO state   in 1D \cite{Kajala:2011},  a nature of two fluids of bound pairs and  free fermions was indicated. 

\begin{figure}[ht]
\includegraphics[width=0.50\textwidth]{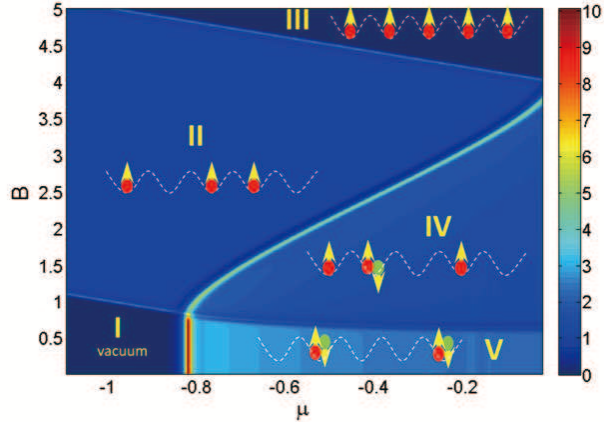}
\caption{Low temperature phase diagram determined by the contour plot of the  Wilson ratio
$R_\textrm{W}^{\kappa}$  (\ref{W-kappa}) calculated from the TBA equations  (\ref{tba1})-(\ref{tba3})  for  the 1D attractive Hubbard model in the  $\mu$-$B$ plane  at  $T=0.01$ and $u=-1$. 
Here the setting is chosen  for better visibility.
Up (down) spins are  represented by red (green) balls. Different values of this ratio uniquely  present five quantum phases.
Sudden enhancement of the ratio in the vicinity of critical lines marks the phase boundaries between different phases, see text.}
\label{fig1:phase-diagram}
\end{figure}

In this Rapid Communication, building on the thermodynamic Bethe ansatz (TBA) equations  of the attractive Hubbard model,
we  analytically obtain the  FFLO pairing correlation and  the universal two free quantum fluids of the FFLO-like state,
where the lattice effects are seen to drive the system differently to the continuous Fermi gas \cite{Yan67,Gau67,Orso:2007,Hu:2007,Lia10,GuaBL13}, see Fig.~\ref{fig1:phase-diagram}.
More detailed  studies of  this  model will be  presented  elsewhere \cite{Cheng:2017-preprint-1,Cheng:2017-preprint-2}.

{\bf The  Bethe ansatz  solution.} The 1D single  band Hubbard model  is described by the Hamiltonian \cite{Ess05}
\begin{align}
H=&- \sum^L_{j=1, a=\uparrow,\downarrow} \left( c_{j,a}^\dagger c_{j+1,a} + {\rm h.c.}   \right)  \notag \\
&+ u \sum_{j=1}^L \left(2n_{j,\uparrow}-1\right) \left(2 n_{j,\downarrow}-1\right), \notag
\end{align}
where $c_{j,a}^\dagger$ and $c_{j,a}$ are the creation and annihilation operators of electrons (fermionic atoms) with
spin $a$ (internal degrees of freedom)  ($a=\uparrow$ or $a=\downarrow$) at site $j$ on  a 1D  lattice with length $L$.
They satisfy the anticommutation relations $\{ c_{j, a}, c_{k,b} \}= \{ c_{j, a}^\dagger , c_{k,b}^\dagger \}=0$ and
$\{ c_{j, a}, c_{k,b}^\dagger \}=\delta_{jk}\delta_{ab}$.
Meanwhile $n_{j,a}=c_{j,a}^\dagger c_{j,a}$ is the density operator, $n_e=\frac{1}{L}\sum_{j=1}^L\sum_{a} n_{j,a}$ is the total fermion number per lattice  site
and $u$ is the dimensionless interaction strength between particles ($u>0$ for  repulsion and $u<0$ for  attraction).

In 1968  Lieb and Wu  \cite{Lieb:1968} derived  the Bethe ansatz (BA) equations for  the 1D Hubbard  model by means  of Bethe's hypothesis \cite{Bethe:1931}.
Takahashi \cite{Takahashi:1972,Takahashi:1974} discovered the solutions of the BA equations which in general are
classified as real quasimomenta $k$, $k$-$\Lambda$ strings and complex spin rapidities of  $\Lambda$ strings, see \cite{Tak99}.
These  roots respectively count for the quasimomenta of the single fermions, bound states of different lengths of fermions and
bound states of magnons with different lengths. At  high energy or momentum, such bound states can  coexist.
Building on Takahashi's string hypothesis, we obtain the TBA equations for the 1D attractive Hubbard model \cite{Cheng:2017-preprint-1}
{\small
\begin{eqnarray}
\varepsilon(k)&=& g_0(k)-\sum_{n=1}^\infty a_n \ast \left( F[\varepsilon^{\prime}_n]-F[\varepsilon_n]\right)(k),\label{tba1} \\
\varepsilon_n(\Lambda)&=&2nB-a_n^t \ast F[\varepsilon](\Lambda)
 -\sum_{m=1}^\infty A_{n m} \ast F[\varepsilon_m](\Lambda), \label{tba2}\\
\varepsilon_n^\prime(\Lambda)&=&g_n(\Lambda) -a_n^t \ast F[\varepsilon](\Lambda)  -\sum_{m=1}^\infty A_{n m} \ast F[\varepsilon^\prime_m](\Lambda) \label{tba3}
\end{eqnarray}
}with  the notation $F[x](y)=-T\ln[1+\exp(-{x(y)}/{T})]$ and $n=1,\ldots,\infty$.
The kernel function $a_n(x)=\frac{1}{2\pi} \frac{2n|u|}{(n|u|)^2+x^2}$.
The driving terms  are $g_0(y)= -2\cos y-\mu-2u-B$ and $g_n(y) =-4\textmd{Re}\sqrt{1-(y+\mathrm{i} \, n \, |u|)^2}-n\left(2\mu+4u\right)$.
In the above equations we denoted the convolutions
$a_n \ast F[x] (k) = \int_{-\infty}^{\infty}
 \textmd{d}y \, a_n( k-y) F[x(y)]$ and $
a_n^t \ast F[x](\Lambda)  =\int_{-\pi}^{\pi}
\textmd{d}  y \, \cos y \,a_n(\sin y-\Lambda) F[x(y)]$.
The functions $\varepsilon$,  $\varepsilon_m^\prime$ and $\varepsilon_n$  stand for the dressed energies  for unpaired fermions,
bound states of $2m$ fermions (the $k$-$\Lambda$ strings) and   length-$n$ spin strings of magnons, respectively.
The function $A_{nm}(x)$ is given in  \cite{Cheng:2017-preprint-1}.

It is particularly important to observe that the  longer $k$-$\Lambda$ strings are  involved in the thermodynamics  as  temperature increases \cite{note-k-L-string}.
The free energy per site is thus given by
\begin{eqnarray}
f=u+\int_{-\pi}^\pi \frac{\texttt{d}k}{2\pi} F[\varepsilon](k)
+\sum_{n=1}^\infty\int_{-\infty}^{\infty}\frac{\texttt{d}\Lambda}{2\pi}
\xi_n(\Lambda) F[\varepsilon_n^\prime](\Lambda)\label{FE}
\end{eqnarray}
with $\xi_n(\Lambda)=\int_{-\pi}^\pi \textmd{d} k
\, a_n(\Lambda-\sin k)$. We also observe that in the dilute limit, $u\to 0$, $n_e\to 0$ with $n_e/|u|$ constant \cite{Krivnov:1975},
the TBA equations (\ref{tba1})-(\ref{tba3}) reduce to those of  the Gaudin-Yang model \cite{Ess05,Tak99,GuaBL13}.
We note that the Shiba transformation between the repulsive and attractive regions of the Hubbard model does not help to obtain universal low energy physics from the TBA equations.
This is mainly because the cut-off processes regarding the above spin and  charge bound states are quite different \cite{Cheng:2017-preprint-1}, unlike the case of  the ground state \cite{Essler:1994}.
As we shall see, in the attractive regime, the low energy physics of the model
 is no longer described by the spin-charge separated theory, rather it is described by the FFLO-like quantum liquids of pairs and single fermions.

{\bf Quantum phase diagram and Wilson ratio.}
 In contrast to the repulsive case, the ground state  of the attractive Hubbard model has  charge bound states, i.e.,
 length-$1$ $k-\Lambda$ strings,  forming  a lattice version of the FFLO state.
 The quantum phases  and phase diagram at $T=0$  can be directly determined   from the  TBA equations   (\ref{tba1})-(\ref{tba3}) in the limit $T\to 0$, which are  called the dressed energy equations \cite{Cheng:2017-preprint-1}.
The dressed energy equations  determine  five quantum phases in the $\mu$-$B$ plane: vacuum I, fully-polarized phase II,
half-filled phase III, FFLO-like state IV and fully-paired state V, see Fig.~\ref{fig1:phase-diagram}.
%
The zero temperature  phase boundaries can also be  determined by  the Shiba transformation \cite{Ess05}.

Here we show that the Wilson ratio, namely, the dimensionless ratio of the  compressibility $\kappa$ and the specific heat divided by the temperature $T$,
\begin{equation}
R_\textrm{W}^{\kappa}=\frac{\pi^2 k_\textrm{B}^2}{3} \frac{\kappa}{C_v/T},\label{W-kappa}
\end{equation}
provides a convenient way for revealing the full phase diagram at low temperatures, see Fig.~\ref{fig1:phase-diagram}.
In the above $k_\textrm{B}$ is Boltzmann's constant.
This  ratio can be directly calculated from the finite temperature TBA equations   (\ref{tba1})-(\ref{tba3}) with a suitable spin and charge bound state cut-off  process,  see \cite{Cheng:2017-preprint-1}.
%
We find that the ratio $R_\textrm{W}^{\kappa}$  is capable of distinguishing all phases of quantum states,
including the FFLO-like state in the phase diagram  Fig.~\ref{fig1:phase-diagram}.
We observe that an enhancement of this ratio  occurs  near a phase transition.
It gives a finite value  at the critical point  unlike the divergent values of compressibility and  susceptibility for $T\to 0$.
 Indeed, the phase boundaries determined  by the Wilson ratio  (\ref{W-kappa}) coincide with the ones determined by  the
 dressed energy equations  at $T=0$.

\begin{figure}[t]
\centering
\includegraphics[width=0.50\textwidth]{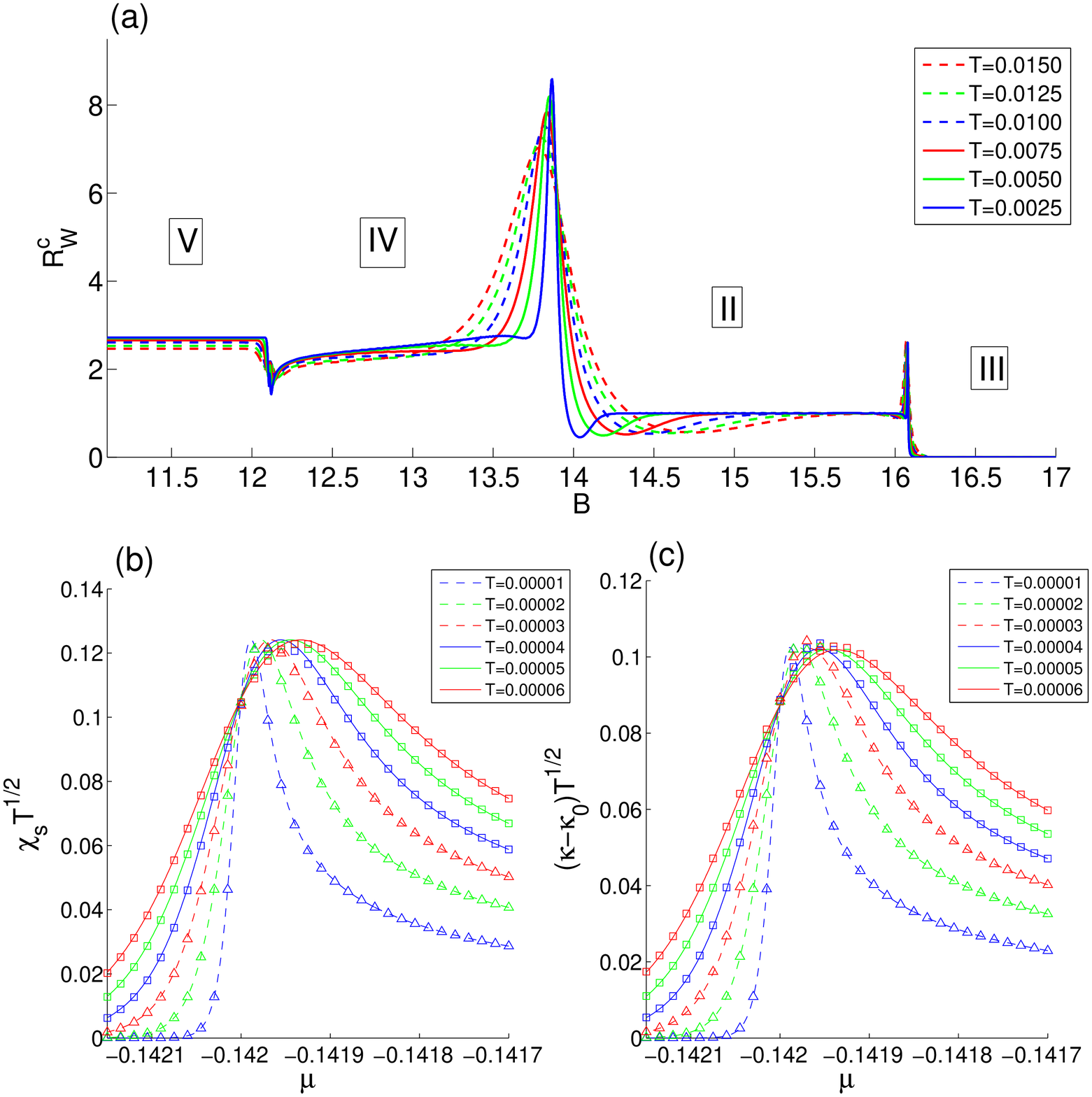}
\caption{(a) Wilson Ratio $R_\textrm{W}^{\kappa_c}$ vs magnetic field for fixed  $\mu=-0.08$ and $u=-7$ in the strong coupling regime.
The sharp peaks at phase transitions  distinguish different quantum  phases V, IV, II and III, respectively.
The constant values of the ratio show Fermi liquid nature in these phases.
(b) and (c) show the scaling invariant behaviour of the susceptibility and compressibility for a fixed  $B=12.142$. The numerical TBA results
(lines) perfectly confirm the analytical  scaling functions (\ref{scale1}) and (\ref{scale2}) (symbols).  }
\label{fig:scaling}
\end{figure}

 The phases IV  and V  in  Fig.~\ref{fig1:phase-diagram}  reveal  significant features, namely  the quasi-long range order  and free-fermion quantum criticality.
 A constant Wilson ratio implies that
the two types of fluctuations are on an equal footing, regardless of the
microscopic details of the underlying many-body systems.
 Regarding the sudden change of the Wilson ratio near a phase transition,  we observe that  the  particle number and energy
 fluctuations become temperature dependent, see Fig.~\ref{fig:scaling}(a).
  At the critical point,  the vanishing of the Fermi points, i.e., $\varepsilon_1(0)=0$ and $\varepsilon_1^\prime (0)=0$,
  in the Fermi sea of pairs and of unpaired fermions leads to a universality class of quantum criticality.
  In the critical regime, the scaling functions of thermodynamic  properties can be cast into  universal forms.
  From the TBA equations   (\ref{tba1})-(\ref{tba3}),  we obtain the scaling functions of compressibility and susceptibility
\begin{eqnarray}
\kappa(\mu,B,T)&=& \kappa_0(\mu,B)+T^{\frac{d}{z}+1-\frac{2}{\nu z}} \lambda_\kappa \mathcal{F}\left( \frac{\mu-\mu_c}{T^{1/\nu z}} \right) \!, \label{scale1}\\
\chi(\mu,B,T)&=& \chi_0(\mu,B)+T^{\frac{d}{z}+1-\frac{2}{\nu z}} \lambda_\chi \mathcal{K}\left( \frac{\mu-\mu_c}{T^{1/\nu z}} \right) \! . \label{scale2}
\end{eqnarray}
Here  the scaling functions $\mathcal{F}(x)=\mathcal{K}(x)= \textrm{Li}_{-1/2}(x)$ indicates a free-fermion criticality classified by the
dynamical critical exponents $z=2$ and correlation critical exponent
$\nu=1/2$, see \cite{Cheng:2017-preprint-1}.
The  terms $ \kappa_0$ and $\chi_0 $ are the regular part and the factors $\lambda _{\kappa,\chi }$ are phase dependent constants.
Fig.~\ref{fig:scaling}(b) and  Fig.~\ref{fig:scaling}(c) show such universal scaling behaviour of the susceptibility and compressibility across the phase boundaries (V,IV).
Similar scaling invariant behaviour occurs whenever the model parameters are driven across the phase boundaries in Fig.~\ref{fig1:phase-diagram}.

{\bf  FFLO correlation.}
For the fully paired state V,  the pairing correlation length is larger than the average interparticle  spacing.
 In this phase, the single particle Green's function decays exponentially, whereas  the singlet pair correlation function decays as a power of
distance \cite{Bogoliubov:1988}.
However,  once the external field exceeds the critical line between phases IV and V, the Cooper pairs start to break apart. Thus both of these
correlation functions decay as a power of distance, indicating a quasi-long range correlation.  In the phase IV, Cooper pairs  and excess fermions form a
1D analogue of the  FFLO pairing-like state \cite{Tezuka:2008,Feiguin:2007}.
However,  analytical result for  the FFLO pairing correlations for the Hubbard model  is still lacking.
 For obtaining a universal  form of   the FFLO-like correlation function, we  first  focus on the  case of  low density $n_e\ll 1$ and low energy.
In the FFLO-like phase IV, the spin wave bound states ferromagnetically couple to  the Fermi sea of  the unpaired fermions.
Thus   the spin wave fluctuations can be ignored at low temperatures due to this ferromagnetic nature.
Then we simplify the TBA equations  (\ref{tba1})-(\ref{tba3})  as  \cite{Cheng:2017-preprint-1}
\small{
\begin{eqnarray}
\varepsilon(k)&\approx &k^2-\mu_1
-a_1 \ast F[\varepsilon^{\prime}_1](k),\label{tba-S1} \\
\varepsilon_1^\prime(\Lambda)&\approx &\alpha_1 \left( \Lambda^2- \mu_2\right)-a_1 \ast F[\varepsilon](\Lambda)
- a_2 \ast F[\varepsilon^\prime_{1}](\Lambda).\label{tba-S2}
\end{eqnarray}
}
The free energy (\ref{FE}) reduces to
$f\approx u+\int_{-\pi}^\pi \frac{\texttt{d}k}{2\pi} F[\varepsilon](k)
+\int_{-\infty}^{\infty}\frac{\texttt{d}\Lambda}{2\pi}
\beta_1 F[\varepsilon_1^\prime](\Lambda)$   \cite{note1}.
In this new set of TBA equations (\ref{tba-S1}) and (\ref{tba-S2})  we have introduced two effective chemical potentials
\begin{eqnarray}
\mu_1&=&\mu-2|u|+B+2,\nonumber\\
\mu_2&=&\frac{1}{\alpha_1 }\left[2\mu+
4(\sqrt{u^2+1}-|u|)\right],\label{CP}
\end{eqnarray}
for understanding the FFLO correlation and  free-fermion nature of  the attractive Hubbard model.
In the above equations the  parameters $\alpha_n$  and $\beta_n$ reflect the interacting effect of the length-$n$ $k$-$\Lambda$ bound states on a lattice. They are given by
\begin{eqnarray}
\alpha_n &=&\int_{-\pi}^\pi \textmd{d} k \,\cos^2 k\,a_n(\sin k)  \,\frac{ 2|u| \cos^2 k (n^2u^2-3 \sin^2 k)}{(n^2u^2+\sin^2 k)^3},\nonumber\\
\beta_n&=& \int_{-\pi}^{\pi} \textmd{d} k \, a_n(\sin k). \nonumber
\end{eqnarray}
At low energy physics only length-$1$ $k$-$\Lambda$ strings  are involved.
In this region, the lattice parameters $\alpha_1$ and $\beta_1$  approach $2$ when $u$ tends to zero.
However, for large $|u|$,  the band of  pairs becomes flat \cite{Cheng:2017-preprint-1}.  The TBA equations (\ref{tba-S1}) and (\ref{tba-S2})
are reminiscent  of the `feedback interaction' equation in the  Landau Fermi liquid theory \cite{Coleman,Wan98}.
The driving term in  (\ref{tba-S2})  can be expressed  as
$\frac{\hbar^2}{2m}\alpha_1(k^2-\mu_2)=\frac{p^2_0}{2\alpha_1 m}-\frac{\hbar^2 }{2m}\alpha_1 \mu_2 $ with $2m=\hbar=1$,
which  is the first-order coefficient describing the excitation energy of a single bound pair.
The lattice parameter  $\alpha_n$ characterizes  the effective mass  of length-$n$ $k$-$\Lambda$ strings (bound state of $2n$ atoms on a lattice).

In light  of the conformal field theory approach  \cite{Bogolyubov:1989,Frahm:1990,Frahm:1991}
and using the TBA equations  (\ref{tba-S1}) and (\ref{tba-S2}),
we calculate  the  asymptotic form  of  the FFLO  correlation function of the attractive Hubbard model in  the low density region \cite{note-asymptotic}
\begin{eqnarray}
G_p(x,t)&=&\langle \Psi_\uparrow^\dag(x,t) \Psi_\downarrow^\dag(x,t) \Psi_\uparrow(0,0) \Psi_\downarrow(0,0) \rangle \nonumber\\
&\approx& A_{p,1} \frac{\cos \left( \pi(n_\uparrow-n_\downarrow)x \right)}{|x+{\rm i} \, v_u \, t|^{2\theta_1}\,|x+{\mathrm i} \, v_b \, t|^{2\theta_2}}\nonumber\\
&&+ A_{p,2} \frac{\cos \left( \pi(n_\uparrow-3n_\downarrow)x \right)}{|x+{\mathrm i} \, v_u \, t|^{2\theta_3}\,|x+{\mathrm i} \, v_b \, t|^{2\theta_4}},
\end{eqnarray}
with the exponents $\theta_1 \approx 1/2, \quad
\theta_2 \approx 1/2+\frac{n_2}{|u|\beta_1} $,  $\theta_3 \approx \frac{1}{2}-\frac{4\, n_2}{|u|\beta_1}$ and
$\theta_4 \approx \frac{5}{2}-\frac{4\,n_1}{|u|}-\frac{3\, n_2}{|u|\beta_1}$. Here $n_{2,1}=N_{2,1}/L$ are  the dimensionless  densities  of pairs and  unpaired fermions, respectively.
The sound velocities are given by $v_b=\frac{\sqrt{\alpha_1 }}{\beta_1}\pi n_2\left(1+\frac{1}{|u|\beta_1 } \left(2n_1+n_2\right)\right)$ and $v_u=\sqrt{2} \pi n_1 \left(1+\frac{4}{|u|}n_2 \right)$.
In the above equation the coefficients $A_{p,1}$ and $A_{p,2}$ are  constant factors. In this phase IV
the spatial oscillation in  the pairing correlation is a characteristic  of the FFLO state, where the
imbalance $n_{\uparrow}-n_{\downarrow}$ in the densities of spin-up and spin-down fermions gives rise to a mismatch in
Fermi surfaces between both species of fermions.
In 1D,  the spatial oscillation signature in pair correlation is  a consequence of  the  backscattering for bound pairs and unpaired fermions, see also the results for the
Gaudin-Yang model \cite{Lee-Guan:2011}.
Here we observe that the critical exponent $\theta_2$ depends essentially on the lattice  parameter $\beta_1$.
So do the critical exponents in other types of correlation functions \cite{Cheng:2017-preprint-2}.
The Fourier transform of  $G_P(x,0^+)$ gives
$\tilde{G}_p(k) \sim  \left[ \textmd{sign}\left(k-\pi(n_\uparrow-n_\downarrow) \right) \right]^{2s_p}|k-\pi(n_\uparrow-n_\downarrow)|^{\nu_p}$
with $2s_P\approx 0$ and $\nu_p \approx n_2/(|u|\beta_1)$.

\begin{figure}[ht]
\centering
\includegraphics[width=0.50\textwidth]{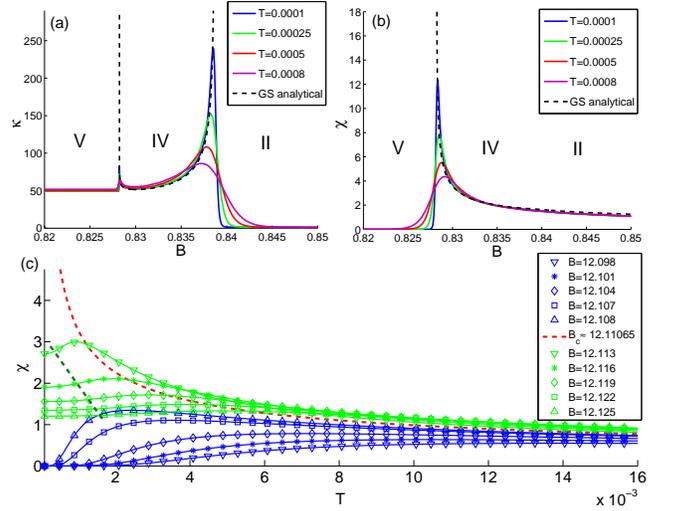}
\caption{(a)  Compressibility $\kappa$ and and (b) spin susceptibility $\chi$  vs magnetic field $B$  for the attractive Hubbard  model with   $u=-1$ and $\mu=-0.8282$. The black dashed lines show the result obtained from the  additivity rules (\ref{Add-Ru}) with the relations  (\ref{relation}) in  Phase IV.
All  compressibility and susceptibility curves at low temperature merge into their zero temperature ones given by the additivity rules  (\ref{Add-Ru}).
 (c) The red dashed line shows the susceptibility at the critical magnetic field $B_c$.
 For $B<B_c$ the susceptibility shows an exponential decay with the energy  gap (\ref{Sus-gap}).
 For $B>B_c$ the susceptibility is almost temperature independent for the gapless phase,
 see the region left of  the green dashed line. Here the parameters are $u=-7$ and $\mu=-0.08$ for the  strong coupling regime. }
\label{fig:sus}
\end{figure}

{\bf Two free fluids  and spin gapped phase.}
At low temperatures, we find  a significant  nature of two fluids in phase IV.
For the ground state, the energy can be regarded as two  TLLs  of unpaired fermions and of pairs due to the quasi-long range correlation.
Without losing generality,  we consider a physical regime  of  low density ($n_e$ small),  low temperature  and finite strong magnetic field.
This region is reachable in cold atoms \cite{Boll:2016}.
In this regime, the chemical potentials for the unpaired fermions and pairs are given explicitly by
\begin{eqnarray}
\mu_1&=& \pi n_1^2A_1^2+\frac{4\pi^2\alpha_1}{3\beta_1^3|u|}n_2^3A_2^3, \label{chemical-1}\\
\mu_2&=&\pi^2\frac{n_2^2}{\beta_1^2}A_2^2+\frac{4\pi^2}{3\alpha_1|u|}n_1^3 A_1^3 + \frac{2\pi^2}{3\beta_1^3|u|}n_2^3 A_2^3 \label{chemical-2},
\end{eqnarray}
where    $A_1=1+\frac{2n_2}{|u|}+\left(\frac{2n_2}{|u|}\right)^2$ and $A_2=1+\frac{2n_1+n_2}{\beta_1 |u|}+\left(\frac{2n_1+n_2}{\beta_1 |u|}\right)^2$  indicate  interacting effects among pairs and unpaired fermions like that of the Fermi gas  \cite{Guan:2007}.
%
The effective chemical potential $\mu_2$ in  the 2D interacting Fermi gases  shows
a  crossover from a Bose-Einstein condensate to a Bardeen-Cooper-Schrieffer superconductor in ultracold fermions \cite{Boett15}.
Moreover, from the relations (\ref{CP}) we   demonstrate    the free-particle nature of two fluids  through  the  additivity rules in compressibility and susceptibility:
\begin{equation}
\kappa=\kappa_1+\frac{2}{\alpha_1 }\kappa_2, \qquad \frac{1}{\chi}=\frac{1}{\chi_1} +\frac{\alpha_1 }{2}\frac{1}{\chi_2},\label{Add-Ru}
\end{equation}
where $\kappa_{r}=\left( \partial  r \,n_{r}/\partial \mu_{r}\right)|_B$ and $\chi_r=\left( \partial r \,n_r/\partial \mu_r \right)|_n$ with $r=1,2$ for unpaired fermions and pairs, respectively.
We see that  the effective binding energy $e_{b}=-(2u+2)n_1-4(u+\sqrt{u^2+1})n_2$ of a bound pair is  absorbed into the effective chemical potentials.
The compressibility and susceptibility can be explicitly calculated from
the chemical potentials (\ref{chemical-1}) and  (\ref{chemical-2}) via the  relations
\begin{eqnarray}
\frac{1}{\kappa_1}&=&\frac{J}{(  \frac{\partial \mu_1}{\partial n_2}
-\frac{\alpha_1}{2}\frac{\partial \mu_2}{\partial n_2} )},\qquad \frac{1}{\kappa_2}=-\frac{1}{\alpha_1}\frac{J}{( \frac{\partial \mu_1}{\partial n_1}
-\frac{\alpha_1}{2}\frac{\partial \mu_2}{\partial n_1})},\nonumber\\
\chi_1&=&\frac{1}{(\frac{\partial \mu_1}{\partial n_1}
-\frac{1}{2}\frac{\partial \mu_1}{\partial n_2}) },\qquad
\chi_2=-\frac{1}{(\frac{\partial \mu_2}{\partial n_1}
-\frac{1}{2}\frac{\partial \mu_2}{\partial n_2}) },\label{relation}
\end{eqnarray}
where the Jacobi determinant
$J=-\frac{\alpha_1}{2}
(\frac{\partial \mu_1}{\partial n_1}\frac{\partial \mu_2}{\partial n_2}
-\frac{\partial \mu_2}{\partial n_1}\frac{\partial \mu_1}{\partial n_2})$.
The explicit forms are given in  \cite{Cheng:2017-preprint-1}.
The additivity rules in the thermodynamic properties reveal a significant free-particle feature in the phase of multiple quantum liquids on a 1D lattice.
Furthermore, using the TBA equations (\ref{tba1})-(\ref{tba3}) and the BA equations with the  length-$1$ $k-\Lambda$ strings,
we show that the specific heat, i.e., a measure  of the energy fluctuations,
is given by $C_v=\frac{\pi T}{3}\left( \frac{1}{v_u} + \frac{1}{v_b}\right)$.
Here the sound velocities $v_{b,u}$ are as given above.

A second-order  phase transition occurs when the system is driven across the phase boundary in the $\mu-B$ plane, see Fig.~\ref{fig1:phase-diagram}.
Fig.~\ref{fig:sus}(a) and Fig.~\ref{fig:sus}(b)  show  the compressibility and susceptibility  vs  magnetic  field at different temperatures.
 They  are temperature independent in  phase IV, whereas the specific heat depends linearly on the temperature, having thus a common feature of the Fermi liquid in higher dimensions.
We observe that in phase IV the compressibility and susceptibility curves at different temperatures collapse into the zero temperature ones   obeying  the additivity rules (\ref{Add-Ru}).
 Fig.~\ref{fig:sus}(c) shows the susceptibility vs  temperature for different magnetic  fields.
For  $B>B_c$, the susceptibility  displays a flat region in the $\chi-T$ plane, the small region  to the left of  the green dashed line, indicating the two free fluids.
For $B<B_c$, the susceptibility illustrates the exponential decay as temperature decreases (blue lines).
In this case, the susceptibility is given by $\chi_s=\frac{T^{-1/2}}{4\sqrt{\pi}}\mathrm{e}^{-\Delta/T}$ with the energy gap
 \begin{equation}
 \Delta=-R^2+\frac{4 (2\pi-R^3/3)}{3 |u| \pi^2} \left( 1+ \frac{2|u|\pi\mu}{2\pi-R^3/3} \right)^{3/2},  \label{Sus-gap}
 \end{equation}
 indicating the behaviour of dilute magnons.
Here we have denoted  $R=\textrm{Re}\sqrt{\mu+2u+B+2}$.

In summary,  for the attractive Hubbard model, we have analytically calculated  the  FFLO pair correlation and critical exponents,  along with
scaling functions of thermal and  magnetic properties for  which the lattice effect becomes prominent.
We have obtained   the effective chemical potentials  of the bound pairs and of the unpaired fermions  and demonstrated  the
additivity rules of  the susceptibility and the compressibility in the FFLO-like state.
While we  have found  that  the susceptibility and the compressibility are temperature independent, the specific heat depends linearly on the temperature in this phase.
These  results provide strong evidence for the existence of two free fluids of bound pairs and of unpaired fermions,
which were predicted in expansion dynamics of the FFLO state in 1D \cite{Kajala:2011}.

{\em Acknowledgments.}  The authors SC and YCY contributed equally to the calculations in this paper.
The authors  thank M. Takahashi and R. Hulet  for helpful discussions. This work is supported by
Key NNSFC grant number 11534014, MOST grant number 2017YFA0304500,
NNSFC grant numbers 11374331, 11174375 and ARC Discovery Projects DP130102839, DP170104934.


\begin{thebibliography}{99}
\bibitem{note-k-L-string} Private communication with Professor M Takahashi. If the effective mass of the $k-\Lambda$ strings is positive, they do not form bound states.
The effective mass is given by $d^2 E(q)/d q ^2$. In the Hubbard model,  the mass can be negative at high energy or momentum. Therefore $k-\Lambda$ string bound states can exist at high energy scales.
\bibitem{note1}We observe that  a small value of $\Lambda$ makes a realistic contribution to the free energy of the pairs in the dilute limit.
Therefore for  low density  only the leading order  of $\xi_n(\Lambda)$, i.e., the  $\beta_n$, is used  for  the free energy of the pairs.
\bibitem{note-asymptotic}Here the asymptotic behavior means the long distance behavior, i.e., the distance $x$ is large.
\end{thebibliography}
\end{document}